\documentclass[12pt]{article}       
\usepackage{latexsym}

\newtheorem{thm}{Theorem}  
\newtheorem{lm}{Lemma} 
\newtheorem{cl}{Corollary}
\newcommand{\bth}{\begin{thm}\hspace{-2mm}{\bf .}\hspace{1.5mm}}        
\newcommand{\eth}{\end{thm}}
\newcommand{\blm}{\begin{lm}\hspace{-2mm}{\bf .}\hspace{1.5mm}}   
\newcommand{\elm}{\end{lm}}
\newcommand{\bcl}{\begin{cl}\hspace{-2mm}{\bf .}\hspace{1.5mm}}   
\newcommand{\ecl}{\end{cl}}
\newcommand{\bpf}{\noindent {\it Proof:} }   
\newcommand{\epf}{\hfill$\Box$\par\vspace{4mm}\noindent}  
\newcommand{\beq}{\begin{equation}} 
\newcommand{\eeq}{\end{equation}\par\noindent}
\newcommand{\beqa}{\begin{eqnarray*}}
\newcommand{\eeqa}{\end{eqnarray*}\par\noindent}
\newcommand{\beqn}{\begin{eqnarray}}
\newcommand{\eeqn}{\end{eqnarray}\par\noindent} 
\def\epi{\to\hspace{-3.5mm}\to}
\def\longepi{\to\hspace{-3.5mm}\to\hspace{-8.5mm}{-}\ }
\def\E{{\cal E}}
\def\T{{\cal T}}
\def\L{L}
\def\op{^{op}}
\def\cause#1#2#3{#1\stackrel{#2}{\leadsto}#3}
\newcounter{count}
\def\referarticle#1#2#3#4#5#6{\refstepcounter{count}\item[\thecount\label{#1}.]{#2}, {\it #3} {\bf #4}, {#5}
{(#6)}.\par\noindent}
\def\referbook#1#2#3#4#5{\refstepcounter{count}\item[\thecount\label{#1}.]{#2}, {\it #3} {#4}
{(#5)}.\par\noindent}
\begin{document}     
  
\pagenumbering{arabic}            
\pagestyle{plain}


\hbox{}   
\par\vskip 1.5 truecm\par
\noindent{\bf QUANTALOIDS DESCRIBING CAUSATION AND PRO-}  
\par\vskip 0.1 truecm\par
\noindent{\bf PAGATION OF PHYSICAL
PROPERTIES}\footnote{Published: {\it Foundations of Physics Letters} {\bf 14}, 133--145 (2001).} 
\par
\vspace{0.631cm} 
\par\parindent=1.6cm {\bf Bob Coecke, David J. Moore and Isar Stubbe}
\par\vskip 0.406 truecm\par  {\it  Dept.\ of
Mathematics, 
Free
University of Brussels,
\par Pleinlaan 2,  B-1050 Brussels, Belgium
\par bocoecke@vub.ac.be
\par\vskip 0.406 truecm\par   Dept.\ of Theoretical Physics,
Universit\'e de Gen\`eve, \par Quai Ernest-Ansermet 24,
CH-1211 Gen\`eve 4, Switzerland\par closcat@hotmail.com
\par\vskip 0.406 truecm\par  and   
\par\vskip 0.406 truecm\par    D\'ept.\ de Math\'ematiques, Universit\'e
Catholique de  Louvain, 
\par Ch$.$ du Cyclotron 2, B-1348 Louvain-la-Neuve, Belgium
\par i.stubbe@agel.ucl.ac.be }
\par\vskip 0.406 truecm\par
\par\vskip 0.406 truecm\par   
\indent  Received 16 October 2000; revised 22 December 2000.
\par\vskip 0.406 truecm\par
\par\vskip 0.406 truecm\par
\noindent 
A general principle of `causal duality' for physical systems, lying
at the base of representation theorems for both compound and evolving
systems, is proved; formally it is encoded in a quantaloidal setting.
Other particular examples of
quantaloids and quantaloidal morphisms appear naturally within this
setting; as in the case of causal duality, they originate from primitive
physical reasonings on the lattices of properties of physical systems.
Furthermore, an essentially
dynamical operational foundation for studying physical systems is
outlined; complementary as it is to the existing static operational
foundation, it leads to the natural axiomatization of `causal duality'
in operational quantum logic.

\par\vskip 0.406 truecm\par
\noindent  Key words: causal duality, property lattice,
galois adjoint, quantaloid.
\par\vskip 0.406 truecm\par
\par\vskip 0.406 truecm\par
\parindent=0.8cm
\noindent{\bf 1. INTRODUCTION}   
\par\vskip 0.406 truecm\par
\par
\noindent  The starting point for our research program is the fact,
already   observed in Eilenberg and Mac$\,$Lane's
seminal paper [\ref{eilML}],  that preordered sets may
be considered as small thin categories.  One  can then
not only reformulate a large part of the theory of
order  structures in categorical terms, but also apply
general categorical  techniques to specific order
theoretic problems.  In particular, the  notion of an
adjunction reduces to that of a residuation [\ref{ML71}]
\S4.5, whereas the notion of a monad reduces to that of
a closure operator
[\ref{ML71}] \S6.1-2.  Now the above categorical notions
have  direct physical interpretations in the context of
axiomatic quantum  theory, the order relation in the
property lattice being semantic  implication and the
meet being operational conjunction 
[\ref{aer82},\ref{aer94},\ref{jaupir},\ref{pir64},\ref{pir76},\ref{pir98}].  In
particular, the  equivalence between suitable
categories of closure spaces and complete  lattices
determined by the existence of monadic comparison
functors  manifests the primitive duality between the
state and property  descriptions of a physical system
[\ref{coemoo},\ref{moo95},\ref{moo97},\ref{moo99}]. Further, this static approach
can be dynamically generalized by  interpreting
morphisms as transition  structures 
[\ref{amcoestu},\ref{coestu99a},\ref{coestu99},\ref{coestu00}], thereby providing an
explicit  physical realization of enrichment.  Finally,
far from being of merely  aesthetic interest, the
categorical approach to operational quantum  theory
allows the recovery of concrete representations of
abstract  notions in the Hilbertian context via the
fundamental theorems of  projective geometry 
[\ref{faufro93},\ref{faufro94},\ref{faufro98}].  In our  opinion,
then, category  theory is as much a tool for the
theoretical  physicist as for the working mathematician.  
     
In this paper we derive a physical principle to which we refer as `causal
duality'. Explicitly, we shall present a common extension of the
representation  theories of deterministic flows  
[\ref{faumoo}] and compound systems 
[\ref{coe99}] to the dual notions of causation and
propagation,  construed as a physical polarity in the
property lattice, and that lifts to a quantaloidal
duality. Further on, we derive other examples of quantaloids that emerge
naturally in
this setting, thus providing an essentially dynamical operational
foundation for studying
physical systems --- the study of such dynamics goes back to
[\ref{moo95}], the approach of which was then conceptualized in
[\ref{amcoestu},\ref{coestu99}] by means of so-called {\it inductions},
and here we go one step further by explicitly imposing (or axiomatizing)
`causal duality'.

Quantales, first introduced in [\ref{mul86}], are complete lattices
$(L,\leq)$ equipped with a not-necessarily commutative binary operation
(a ``multiplication'' of elements of $L$) that distributes on both sides
over arbitrary suprema; a frame is a quantale in which the
``multiplication'' coincides with the binary infimum. A quantaloid is a
category $\underbar{{\rm Q}}$ in which every hom-set is a complete
lattice and where composition of morphisms distributes on both sides
over arbitrary suprema of morphisms; thus a quantale is precisely a
quantaloid with one object. The pertinent functors between two
quantaloids, called quantaloidal morphisms, are of course those functors
that preserve suprema of morphisms. For a survey of the theory of
quantales and quantaloids we refer to
[\ref{pas00},\ref{ros90},\ref{ros96}]; for a quick introduction to
(enriched) category theory -- of which the theory of quantaloids is an
instance -- we refer to [\ref{borstu}]; as minimal preliminaries to this
paper the appendices of [\ref{amcoestu},\ref{coestu99}] may suffice.   

Let us now recall some of the basic notions of the
operational   approach to physics
[\ref{aer82},\ref{pir76},\ref{pir98}], and fix some notation.  
Given a (well defined) {\it physical system} $\Xi$ we
define a {\it  test}
$\alpha$ as a real experimental procedure relative to
the system  where we have defined in advance the
so-called positive response.   We call  such a test
{\it certain} for a particular  realization of $\Xi$
iff we {\it would} obtain the positive response  {\it
should} we perform the experiment.  For two such tests
$\alpha$  and $\beta$ relative to $\Xi$, we set
$\alpha\preceq\beta$ iff $\beta$  is certain whenever
$\alpha$ is certain.  The expression 
$\alpha\preceq\beta$ then reveals a {\it physical law} for
$\Xi$, and of  course $\preceq$ defines a preorder on
${\cal T}(\Xi)$, the collection  of all possible tests
for $\Xi$.  By standard quotienting techniques  we can
now work with the collection $L(\Xi)$ of equivalence
classes of  tests, two tests
$\alpha$ and $\beta$ being equivalent iff both 
$\alpha\preceq\beta$ and $\beta\preceq\alpha$ (which will be denoted as
$\alpha\approx\beta$), that then comes equipped  with a partial order $\leq$ derived
from the preorder $\preceq$.  The  key point of this
setup is that to any such equivalence class 
$[\alpha]$ corresponds an ``element of physical
reality''
[\ref{EPR}],  called a {\it property} of $\Xi$
[\ref{pir76}].  If a test $\alpha$ is  certain  for a
particular realization of the physical system $\Xi$, 
then the corresponding property
$a=[\alpha]$ is said to be {\it  actual} for this
realization --- otherwise it is  {\it potential}. 
Under the (working) hypothesis that
$L(\Xi)$  constitutes a set -- although in principle
all the following holds  also for it being a thin
category -- it can then be proved that 
$(L(\Xi),\leq)$ is a complete lattice: given a
subcollection
${\cal  A}$ of
${\cal T}(\Xi)$, defining $\prod{\cal A}$ as ``choose
any 
$\alpha$ in ${\cal A}$ as you wish and effectuate it'',
provides 
$L(\Xi)$ with a meet induced by
$\bigwedge\{[\alpha]|\alpha\in{\cal  A}\}=[\prod{\cal
A}]$.  By its construction, this meet is a physical 
{\it conjunction} (an aspect to which we will refer to
as [{\sf con}])  --- but the corresponding join has no
{\it a  priori} physical significance, so it cannot be
treated as a  disjunction, {\it e.g.} orthodox quantum
mechanics.  It also clearly  follows that $a\leq b$ in
$L(\Xi)$ can be treated as an {\it  implication
relation} (referred to as [{\sf imp}]), where we say
that 
$a$ is {\it stronger} than $b$.  Finally, for each
particular   realization of a system $\Xi$, we can write
$\varepsilon$ for the  subset of $L(\Xi)$ that contains
precisely all the properties that are  actual for this
particular realization. As any such $\varepsilon$  is a
complete co-ideal, {\it i.e.}, closed under meets [{\sf
con}] and  upperbounds [{\sf imp}], it can be
characterized by its strongest  element
$p_{\varepsilon}=\bigwedge\varepsilon\in L(\Xi)$. 
Therefore, for  each realization of $\Xi$ there exists
a strongest actual property,  which is appropriately
called {\it state} of the system [\ref{pir76}].
\par\vskip 0.406 truecm\par
\par\vskip 0.406 truecm\par
\noindent  {\bf 2. CATEGORICAL DUALITY INDUCED BY
CAUSALITY}
\par\vskip 0.406 truecm\par
\par 
\noindent    
In this section we aim to give a common extension of the
operational theory of, on the one hand, deterministic
flows
[\ref{faumoo}] and, on the other, compound systems
[\ref{coe99}]; the result  of our analysis will be that
the deeper structural ingredient in both situations is
that ``causation is adjoint to propagation''. 

Considering an evolving physical system $\Xi$, any test
$\alpha$   relative to the system at time $t_2$ defines
a test
$\phi(\alpha)$  relative to the system at an earlier
time
$t_1$ as ``evolve $\Xi$ from 
$t_1$ to $t_2$ and effectuate $\alpha$''.  The property 
$[\phi(\alpha)]$ has a clear interpretation, namely
``guaranteeing actuality of
$[\alpha]$''.  The assignment 
${L(\Xi)\to L(\Xi)}:{[\alpha]\mapsto[\phi(\alpha)]}$,
as we will see  below, describes the evolution of
$\Xi$.  On the other hand,  considering two interacting
physical systems
$\Xi_1$ and $\Xi_2$, any  test $\alpha_2$ on
$\Xi_2$ defines a test $\phi(\alpha_2)$ on $\Xi_1$  by
``let the systems interact and effectuate
$\alpha_2$''.  The  assignment
$L(\Xi_2)\to
L(\Xi_1):[\alpha_2]\mapsto[\phi(\alpha_2)]$  now
encodes the interaction of $\Xi_1$ on $\Xi_2$.

Keeping these two cases in mind, for any two property
lattices
$L_1$ and
$L_2$ we dispose of a {\it causal relation}
$\,\leadsto\ 
\subseteq L_1\times L_2$ where: \beqn a_1\leadsto a_2\ 
\mbox{$\Leftrightarrow$ ``actuality of $a_1$ guarantees
actuality of 
$a_2$''.} \eeqn
\blm\label{lemmacausalrelation} By the operational
significance of
$\,\leadsto\,$ the  following holds:
\beqn\label{eqleadsto1} b_1\leq a_1, a_1\leadsto a_2,
a_2\leq b_2&\Rightarrow&b_1\leadsto b_2\\
\label{eqleadsto2}
\forall a_2\in A_2: a_1\leadsto
a_2&\Rightarrow&a_1\leadsto\bigwedge A_2
\eeqn where $A_2$ is a non-empty subset of $L_2$. 
\elm
\bpf  The proof of eq.(\ref{eqleadsto1}) relies on [{\sf
imp}],  eq.(\ref{eqleadsto2}) follows by [{\sf con}].
\epf  (From an axiomatic point of view the conditions
in the previous lemma are `axioms' for a causal
relation.) 

Now consider the following map
prescription:
\beqn\label{propagation} f^*:L_1\setminus K\to
L_2:a_1\mapsto\bigwedge\{a_2\in L_2\mid a_{1}\leadsto
a_{2}\} 
\eeqn with $K=\{a_1\in L_1|\ /\hspace{-1.7mm}\exists
a_2\in L_2:a_1\leadsto a_2\}$, as such avoiding
non-empty meets.
\blm\label{lemmapropagation}  By lemma
\ref{lemmacausalrelation} and the explicit definition
of $f^*$ we have:
\beqn  
\label{isotone1} a_{1}\leq
a'_{1}&\Rightarrow&f^*(a_{1})\leq f^*(a'_{1})\\ 
\label{adjoint1} a_1\leadsto
a_{2}&\Leftrightarrow&f^*(a_{1})\leq a_{2}
\eeqn    where it is understood that $a_1'\not\in K$.
\elm 
\bpf For eq.(\ref{isotone1}), remark that $a_1'\not\in
K\Rightarrow a_1\not\in K$ by lemma
\ref{lemmacausalrelation}, so both
$f^*(a_{1})$ and
$f^*(a'_{1})$ are defined; then computation shows that
indeed
$f^*(a_{1})\leq f^*(a'_{1})$. In eq.(\ref{adjoint1}), the
sufficiency is trivial; to prove necessity is, by [{\sf
imp}], to prove that
$a_1\leadsto f^*(a_1)$, which is true by [{\sf con}]. 
\epf Now it is clear that
$f^*(a_{1})$ is the strongest property of $L_{2}$ the
actuality of which is guaranteed by the actuality of
$a_{1}$, i.e., $f^*$ describes the {\it propagation of
(strongest actual) properties}. Next, set:
\beqn f_*:L_2\to L_1:a_2\mapsto\bigvee\{a_1\in L_1\mid
a_{1}\leadsto a_{2}\}.
\eeqn
\blm\label{lemmacausalassignment}  By
eq.(\ref{eqleadsto1}) and the explicit definition of
$f_*$ we have:
\beqn  a_{2}\leq a'_{2}&\Rightarrow&f_*(a_{2})\leq
f_*(a'_{2})\\
\label{adjoint2} a_1\leadsto a_{2}&\Rightarrow&a_{1}\leq
f_*(a_{2})
\eeqn    
\elm 
\bpf By computation.
\epf If moreover the condition $1_1\leadsto 1_2$ can be
derived  from the physical particularity of the system
under consideration (or formally, if it is an `axiom' on
$\leadsto$), then
$K=\emptyset$ in eq.(\ref{propagation}), and thus:
\beqn f^*(a_{1})\leq a_{2}\Leftrightarrow a_1\leadsto
a_{2}\Rightarrow a_{1}\leq f_*(a_{2}) \eeqn  so it
remains to show that eq.(\ref{IsarFreakout}) below is valid
to obtain adjointness of $f^*$ and $f_*$.
\blm\label{lemmacausalassignmentbis}   By the
operational significance of
$f_*$ (via that of $\leadsto$) we have:
\beqn\label{IsarFreakout} a_{1}\leq
f_*(a_{2})&\Rightarrow& a_1\leadsto  a_{2}.
\eeqn
\elm
\bpf  Since $[\phi(\alpha_2)]\leadsto[\alpha_2]$ by the
definition of
$\phi$, and since
$a_1\leadsto a_2$ implies that $a_1\leq
[\phi(\alpha_2)]$ we obtain that
$f_*([\alpha_2])=[\phi(\alpha_2)]$.  Since
eq.(\ref{IsarFreakout}) is equivalent to 
$f_*(a_{2})\leadsto a_{2}$ this completes the proof.
\epf  (Note that formally eq.(\ref{IsarFreakout}) is an
additional `axiom' on  
$\,\leadsto\,$.)  Physically, lemma 
\ref{lemmacausalassignmentbis} states that there exists
a  well defined ``weakest cause'' $f_*(a_2)$ in $L_1$
of any
$a_2$ in 
$L_2$, so $f_*$ describes the {\it  assignment of
(weakest) causes (for actuality)}. We can now read that
$f^*$ is left adjoint to $f_*$ (denoted as $f^*\dashv f_*$) or, in words, that
``propagation is adjoint to causation''. By general theory on adjoint pairs of morphisms
(see for example [\ref{coemoo}]) we have the following.
\bcl\label{joinmeet}
The propagation $f^*:L_1\to L_2$ is a join preserving map whereas the causation
$f_*:L_2\to L_1$ is a meet preserving map.
\ecl     

In case that $1_1\not\leadsto 1_2$, one can always
extend the domain and codomain of
$f^*$ and
$f_*$ to the upper pointed extensions  
$L_1\dot\cup\underline{1}$ and
$L_2\dot\cup\underline{1}$ of
$L_1$ and
$L_2$, obtained by freely adjoining a new ``top'' element, and then put 
$f_*(\underline{1})=\underline{1}$,
$f^*(\underline{1})=\underline{1}$ and 
$\forall a_1\in K:f^*(a_1)=\underline{1}$. Physically, an
interpretation of
$\underline{1}$ follows from that of $1_1$ and
$1_2$, respectively being existence of $\Xi_1$ and
$\Xi_2$; see also [\ref{sou}]. From a technical point of view this situation
is now not any different from the case discussed above, and so we obtain again
that $f^*\dashv f_*$.
Henceforth we develop the case $1_1\leadsto 1_2$; it is understood
that in the examples where $1_1\not\leadsto 1_2$ we have freely adjoined a new ``top'' so
as to reduce this case to the former by the procedure outlined above.

When considering three property lattices
$L_1$,
$L_2$ and 
$L_3$ and respective
propagations of properties $f^*_{1,2}:L_1\to L_2$ and $f^*_{2,3}:L_2\to L_3$, what can be
said about $f^*_{1,3}:L_1\to L_3$? 
\blm\label{compo}
With obvious notations $f^*_{i,j}\dashv f_*^{i,j}$, we have 
by the operational significance of the causations $f^{i,j}_*$ that
$f_*^{1,3}=f_*^{1,2}\circ f_*^{2,3}$.
\elm       
\bpf  Denoting the corresponding tests for
$f_*^{i,j}(a_j)$ by
$\phi_{i,j}(\alpha_j)$ we clearly have
$\phi_{1,3}(\alpha_3)=\phi_{1,2}(\phi_{2,3}(\alpha_3))$,
so 
$[\phi_{1,3}(\alpha_3)]=[\phi_{1,2}(\phi_{2,3}(\alpha_3))]$
and thus it follows that
$f_*^{1,3}(a_3)=f_*^{1,2}([\phi_{2,3}(\alpha_3)])=
f_*^{1,2}(f_*^{2,3}(a_3))$ for all
$a_1\in L_1$.    
\epf  
Because adjunctions compose -- that is, if $f^*\dashv
f_*$ and $g^*\dashv g_*$ then also $f^*\circ g^*\dashv g_*\circ f_*$ -- we
also obtain
$f^*_{1,3}=f^*_{2,3}\circ f^*_{1,2}$. 
We can read off that the composition of causations stands for ``chaining causal
assignments'' of properties, whereas the composition of propagations then must stand for
``consecutive propagation'' of properties. 

More technically speaking, from corollary
\ref{joinmeet} and lemma \ref{compo} it is now obvious that the property lattices and the
causations organize themselves in (a subcategory of) $\underbar{MCLat}$, and the same
property lattices equipped with the propagations organize themselves in (a subcategory of)
$\underbar{JCLat}$ --- where $\underbar{MCLat}$ (resp. $\underbar{JCLat}$) is the category
of complete lattices and meet preserving (resp. join preserving) maps. As is well-known,
the assignment of adjoints as in
\beqn
\underbar{JCLat}(L_1,L_2)\to\underbar{MCLat}(L_2,L_1):f\mapsto f_*
\eeqn
is an anti-isomorphism (a ``duality'') between the complete lattices of respectively join
and meet preserving maps, ordered pointwisely. In particular, the
conjunction of properties is
``lifted'' to the ``conjunction for causal
assignments'' in the hom-sets of $\underbar{MCLat}$ and, dually, the superposition of
properties is ``lifted'' to the ``superposition of propagations'' in the hom-sets of
$\underbar{JCLat}$.  Rewritten more conveniently, this gives
\beqn
\underbar{MCLat}^{\sf
coop}(L_1,L_2)\cong\underbar{MCLat}(L_2,L_1)^{\sf op}\cong\underbar{JCLat}(L_1,L_2).
\eeqn
Since both $\underbar{MCLat}^{\sf coop}$ and $\underbar{JCLat}$ are quantaloids for the
pointwise ordering of their hom-sets, and because adjoints
compose, we have a representation of our setting in the category
of quantaloids and quantaloid morphisms, denoted as $\underbar{QUANT}$:
\beqn
\underbar{MCLat}^{\sf coop}\stackrel{\sim}{\longleftrightarrow}\underbar{JCLat}
\eeqn
We can conclude all this by:
\bth Causal assignment and propagation of properties are
dualized by a quantaloidal isomorphism {\rm
$F:\underbar{MCLat}^{\sf coop}\stackrel{\sim}{\longrightarrow}\underbar{JCLat}$}. 
\eth 

We will now briefly discuss some examples of this
general setting. The adjunction 
\beqn
[f^*:0_1\mapsto 0_2,\
\mbox{rest}\mapsto 1_2]\dashv[f_*:1_2\mapsto 1_1,\
\mbox{rest}\mapsto  0_1]
\eeqn 
describes  `separation' of the
systems described by $L_1$ and $L_2$, a situation that
previously could not be described in a consistent way
within  quantum theory
[\ref{aer82},\ref{coe99}].  By way of contrast, for 
$L_1$ and $L_2$ atomistic the maps that send atoms to
atoms or the  bottom represent the strongest types of
interaction, or analogously,  maximally deterministic
evolution.  When considering lattices of  closed
subspaces of Hilbert spaces this setting yields 
representational theorems for the description of
compound quantum  systems by the Hilbert space tensor
product
[\ref{coe99}] and description of evolution by
Schr\"odinger flows
[\ref{faumoo}], so it is exactly the enrichment that
allows a joint  consideration of the types of
entanglement encountered in classical  and quantum
physics.
\par\vskip 0.406 truecm\par
\par\vskip 0.406 truecm\par
\noindent   {\bf 3. PHYSICAL ORIGIN OF
CATEGORICAL CONCEPTS}
\par\vskip 0.406 truecm\par
\par 
\noindent
{\it Action of inductions on
properties}
\par\smallskip\par\noindent
With the consideration of the map $\phi$ in section 2 we've introduced a new ``dynamical
ingredient'' to say something more about the physical system than a merely static
description could ever do. This can be pushed even further: one can
trace the origins of causal duality, i.e. ``propagation is adjoint to causation'', back to
a dynamic operational foundation complementary to the static operational foundation. To
that end, one uses as counterpart of the operational notion of test that of ``induction''
[\ref{amcoestu}]. Whereas giving a fully detailed
exposition of this development would lead us too far, we
still think that it is useful to at least outline the most
important ideas; we plan to dig deeper into this matter in a future work.

By an {\it induction} on a physical system $\Xi$ is meant an
externally imposed change of $\Xi$. Such an induction can as such for example be an
imposed evolution or measurement, or, the action of a system on another in case of
so called entanglement.  The collection
${\cal E}(\Xi)$ of all inductions on the system
$\Xi$ (also written
$\E$ if no confusion is possible) is naturally equipped with two
operations$\,$:
\begin{itemize}
\itemsep=-2pt
\item[1.] for $e_1,e_2\in{\cal E}$ two inductions, $e_1\&e_2$ is
the induction that consists of effectuating first $e_1$ and second
$e_2$;
\item[2.] for $\{e_i\mid i\in I\}\subseteq {\cal E}$ with $I$ a
set, $\bigvee_ie_i$ is the induction that consists of effectuating an
arbitrarily chosen element of
$\{e_i\mid i\in I\}$.
\end{itemize}
When focussing on an induction's action on the physical system rather than the
physical procedure associated to such an induction, it seems reasonable to suppose that
$\E$ is a set; then these operations give
$\E$ the structure of a quantale, for clearly we have that $\&$ acts as a product that
distributes on both sides over the join $\bigvee$ (whence the notations). The unit for the
multiplication can be thought of as the induction {\it
freeze}, denoted as $*$.

The crucial link between inductions and tests is now
given by an action of the former on the latter. Namely, for any
$e\in\E$ and
$\alpha\in\T$, we define a ``multiplication'' $e\cdot \alpha\in\T$ as
follows:
\begin{itemize}
\item[] $e\cdot\alpha$ is the test consisting of ``first
executing the induction $e$ and then performing the test $\alpha$'',
the outcome of the test $e\cdot\alpha$ being the one thus obtained for
$\alpha$.
\end{itemize}
Then the operationality of the notions involved assures that:
\begin{itemize}
\itemsep=-2pt
\item[(o)] $\alpha\preceq\beta$ in $\T$ implies that
$e\cdot\alpha\preceq e\cdot\beta$
\item[(i)] $*\cdot\alpha\approx\alpha$
\item[(ii)] $e\cdot(\Pi_i\alpha_i)\approx\Pi_i(e\cdot\alpha_i)$
\item[(iii)] $(\bigvee_ie_i)\cdot\alpha\approx\Pi_i(e_i\cdot\alpha)$
\item[(iv)]
$e_1\cdot(e_2\cdot\alpha)\approx(e_1\& e_2)\cdot\alpha$
\end{itemize}  where $\approx$ denotes the equivalence relation
derived from the preorder $\preceq$ on $\T$, and the products $\Pi$
in (i) and (ii) are product tests.
When reducing the class $\T$ of tests to the
set $\L$ of properties, by quotienting
$\T\epi\L:\alpha\mapsto[\alpha]$, this
``multiplication'' boils down to an action
\beqn\label{action}
\E\times\L\op\to\L\op:(e,a:=[\alpha])\mapsto e\cdot a\
:=[e\cdot\alpha]
\eeqn such that $L^{\sf op}$ exhibits itself as a module of
the monoid
$\E$ in the monoidal category $\underbar{JCLat}$ (for the theory of quantale modules,
consult for instance [\ref{ros90}]).
 
At this point it is handy to re-introduce the
``causal relation'' of  the previous section, albeit adapted to this situation in which we
want to consider many inductions at once: for
$\alpha,\beta\in\T$ and
$e\in\E$, we
put 
\beqn\label{causalrelation}
\cause{\alpha}{e}{\beta}\Longleftrightarrow\alpha\preceq
e\cdot\beta
\eeqn
and with a slight abuse of notation we will also use
$\cause{a}{e}{b}$ for $a,b\in\L$. The latter means thus precisely
that the actuality of the property $a$ before the induction $e$
guarantees the actuality of $b$ after the induction. It is now a
consequence that to any $e\in\E$ one can associate two mappings,
\beqn
e_*:\L\to\L:a\mapsto e\cdot a \ \ \ \ \ \ \ \ \ \\
e^*:\L\to\L:a\mapsto\bigwedge\{b\in\L\mid\cause{a}{e}{b}\}
\eeqn
for which it is clear that $a\leq
e_*(b)\Longleftrightarrow\cause{a}{e}{b}\Longleftrightarrow
e^*(a)\leq b$ and therefore $e^*\dashv e_*:\L\to\L$. Furthermore, the unital
quantale structure that
$\E_*=\{e_*\mid e\in\E\}\subseteq\underbar{MCLat}(\L,\L)$ is naturally
endowed with, corresponds to the one suggested by $\E$; in
particular $(e\& f)_*=e_*\circ f_*$,
$(\bigvee_ie_i)_*=\bigvee_i(e_i)_*$ and $*_*=id_L$. Likewise for the
evident $\E^*\subseteq\underbar{JCLat}(\L,\L)$. When considering both as one
object quantaloids, it is true that $\E^*\cong\E_*^{\sf
coop}$, and as such we recover in this setting the causal
duality, as formalized by the theorem in the previous section.
\par\vskip 0.406 truecm\par
\par 
\noindent
{\it Operational resolutions and
quantaloids}
\par\smallskip\par\noindent
In order to introduce aspects of `uncertainty' and
`arbitrary choice' in our general setup, we want to extend a property lattice
$(L,\bigwedge)$ by introducing `propositions' that
represent disjunctions of properties [{\sf dis}]. We may 
realize this within
$PL:=2^{L\setminus\{0\}}$ {\it grosso modo} as follows: The embedding
of the lattice
$L$ of properties into the boolean algebra $PL$ of propositions,
\beqn\label{embedding}
L\hookrightarrow PL:x\mapsto\{y\in L\setminus\{0\}\mid y\leq x\},
\eeqn 
preserves arbitrary infima
such that [{\sf con}] and [{\sf imp}] are
preserved. Consequently, the embedding has a left adjoint which turns out to
be
\beqn\label{resolution}
PL\to L:A\mapsto\bigvee A.
\eeqn  
Such a map was dubbed `operational resolution' in
[\ref{amcoestu},\ref{coestu99a},\ref{coestu99}], for the following reason:
this map physically stands for the {\it
verifiability} of collections of properties
in the sense that, if we define an `actuality set' to be an $A\subseteq L$ of
which at least one element $a\in A$ is actual but we don't know which one, then by $\bigvee
A=\bigwedge\{b\in L\mid\forall a\in A:b\geq a\}$, [{\sf con}] and [{\sf imp}],
we have that
$\bigvee A$ is the strongest property whose actuality is
guaranteed for an actuality set $A$. By its construction, it is clear that
in the ambient boolean algebra $PL$ an actuality set $A$ plays the role of
the `disjunction' of its elements.

How should one now describe the propagation of actuality sets? The
answer to this question is given by the more general results in [\ref{coestu99}];
here is what it comes down to:
\blm\label{propactset}
Given two lattices $L_1$ and $L_2$, and a map $g\colon PL_1\to
PL_2$ that preserves arbitrary unions, the following are equivalent:
\begin{itemize}
\itemsep=-10pt
\item[1.] for all $A,B\in PL_1$, 
\beqn\label{continuity}
\bigvee A=\bigvee
B\Rightarrow\bigvee g(A)=\bigvee g(B)\eeqn
\item[2.] there exists a (necessarily unique) morphism $f:L_1\to L_2$ that
preserves arbitrary suprema making the following square, in which the
vertical uparrows are the resolutions, commute:
\[\begin{array}{ccc}
L_1 & \stackrel{f}{\longrightarrow} & L_2 \\
\uparrow & & \uparrow \\
PL_1 & \stackrel{g}{\longrightarrow} & PL_2
\end{array}\]
\end{itemize}
\elm
Indeed, such maps $g:PL_1\to PL_2$ are the appropriate expressions for the
propagation of actuality sets: requiring $g$ to preserve arbitrary unions is
in accordance with [{\sf dis}], and requiring the continuity condition of
eq.(\ref{continuity}) is, via lemma \ref{propactset}, saying that the verification of the
propagation of an actuality set through the operational resolution must result in a
propagation of properties.

Further application of the general theory of resolutions
and their morphisms [\ref{coestu99}] shows that, when defining the
bicategory $Q\underbar{$^{\#}$JCLat}$ with objects the
$PL=2^{L\setminus \{0\}}$ for all complete lattices
$L$, morphisms the maps
$g$ as above, and the local structure being the
evident pointwise order of such maps, this bicategory is a
quantaloid; furthermore, the action 
$F_\#:Q\underbar{$^{\#}$JCLat}\to\underbar{JCLat}:   
PL\mapsto L; g\mapsto f$ (notation of $f$ refers to lemma
\ref{propactset}) proves to be a full quantaloidal morphism. Note
that 
$Q\underbar{$^{\#}$JCLat}$ neither coincides with the
categories with the same objects and on the one hand all
union preserving maps, and on the other hand pointwise
unions of direct image maps of $\bigvee$-preserving maps
--- a precise characterization can be found in
[\ref{coestu00}]. Together with the theorem of section 2 we have the following scheme in
$\underbar{QUANT}$:
\beqn   
Q\underbar{$^{\#}$JCLat}\ 
\stackrel{F_\#}{\longepi}\ 
\underbar{JCLat}
\stackrel{\cong}{\longleftrightarrow}
\underbar{MCLat}^{\sf coop} 
\eeqn  
that expresses how the propagation of
actuality sets is related to causal assignments.  

Our point
now is that the quantaloidal nature of
$F_\#$ reveals that the enrichment of the
collection of causal assignments
$\underbar{MCLat}^{\sf coop}$ originates -- physically --
from the presence of an underlying uncertainty encoded
in the local structure of
$Q\underbar{$^{\#}$JCLat}$.
\par\vskip 0.406 truecm\par
\par   
\noindent
{\it Actuality sets and frame
completions}
\par\smallskip\par\noindent
Surely the formal disjunction of properties $a,b,c,...\in L$ may be expressed in
the complete boolean algebra $PL$ of propositions as their union $\{a,b,c,...\}\in PL$;
and consequently the disjunction of $a,b,c,...$ is actual iff at least one of them is ---
which could indicate that the `calculus of actuality sets' as the appropriate `logic of
the propositions' is encoded in $PL$. That the properties can be, on the one hand,
embedded in the propositions without loss of conjunctivity (a primitive notion!) and, on
the other hand, be `recuperated' from them by an operational resolution is a
confirmation of these ideas.  However, the meets in $PL$ of elements that do not
represent properties are in no way to be seen as conjunctions: for $a<b$ we have
$\{a\}\cap\{b\}=\emptyset$ where the conjunction
$\{a\}$ {\it and}\, $\{b\}$ clearly is $\{a\}$.
  
It turns out that the origin of this lack of general conjunctivity  traces back to the
fact that the inclusion
$L\hookrightarrow PL$ is in general not the most ``economical'' way of
extending
$L$ to be able to handle those actuality sets that are necessary to express disjunction of
properties. Indeed, consider the case where
$L$ is already a complete boolean algebra: the extension of
$L\hookrightarrow PL$ is redundant whenever the join
in $L$ is already to be understood  as a disjunction.
For certain classes of property lattices -- among which all physically
relevant ones -- a ``most economical extension'', i.e. a completion
which is universal in an appropriate ambient category, does exist; the
construction is subtle but straightforward. A profound discussion can be found in
[\ref{coeSL}] and we will not go into details here; let us however quickly sketch the
crucial point.
    
It is necessary to somehow characterize, for a given property lattice $L$,
those joins of subsets of $L$ that are disjunctions in the sense that the join is actual
iff at least one member in this subset is. That is to say, we need a lattice theoretic
criterion to decide whether, for example, a binary join 
$a\vee b$ in
$L$ is to be understood as disjunction of
$a$ and $b$ or as superposition (the criterion should work also for arbitrary joins). The
following result, quoted from [\ref{coeSL}], provides such a criterion.
\blm\label{distrjoin}
If the property lattice $L$ ``fully represents the physical system with respect to
superpositions'', the join of a subset
$A\subseteq L$ is to be understood as the disjunction of $A$ iff $\bigvee A$ is
distributive --- that is, for all $x\in L$: $x\wedge
\bigvee A=\bigvee(x\wedge A)$.
\elm 
This makes at once clear that the `logic of propositions' (or the `calculus of actuality
sets') must take place in a complete lattice in which every join is distributive --
because we want the join of propositions to be their disjunction -- hence by definition
in a frame.  So a frame completion of $L$ is what we're looking for. A detailed discussion --
with appropriate references -- of such completions and their validity,
consequences and applications in the field of quantum logic, and alternative constructions can
be found in [\ref{coeSL},\ref{stu}].  
\par\vskip 0.406 truecm\par
\par\vskip 0.406 truecm\par
\noindent   {\bf REFERENCES}
\begin{itemize}    
\itemsep=-8pt  
\itemindent=-0pt            
\labelsep=2pt  
\leftskip=-29pt
\referarticle{aer82}{D. Aerts}{Found.~Phys.}{12}{1131}{1982}
\referarticle{aer94}{D. Aerts}{Found.~Phys.}{24}{1227}{1994}
\referarticle{amcoestu}{H. Amira, B. Coecke, and I.
Stubbe}{Helv.~Phys.~Acta}{71}{554}{1998}
\referarticle{birneu}{G. Birkhoff and J. von Neumann}{Ann.~Math.}{37}{823}{1936}
\referbook{borstu}{F. Borceux and I. Stubbe}{Short Introduction to Enriched
Categories,}{in: Current Research in Operational Quantum Logic, edited by B. C\mbox{oe}cke, D.J.
Moore and A. Wilce, pp.167--194}{Kluwer Academic Publishers, Dordrecht, 2000}
\referarticle{coe99}{B. Coecke}{Int.~J.~Theor.~Phys.} {39}{581}{2000}
\referarticle{coeSL}{B. Coecke}{{\rm Quantum logic in intuitionistic perspective \&
disjunctive quantum logic in dynamic perspective, $\langle$http://xxx.lanl.gov$\rangle$
arXiv: Math.LO/0011208
\& 0011209}}{$\!\!$}{{\rm to appear in} {\it Studia Logica}}{2001}  
\referbook{coemoo}{B. Coecke and D.J. Moore}{Operational Galois adjunctions,}{in: Current
Research in Operational Quantum Logic, edited by B. Coecke, D.J. Moore and A. Wilce,
pp.195--218}{Kluwer Academic Publishers, Dordrecht, 2000}
\referarticle{coestu99a}{B. Coecke and I. Stubbe}{Int.~J.~Theor.~Phys.}{38}{3296}{1999}
\referarticle{coestu99}{B. Coecke and I. Stubbe}{Found.~Phys.~Lett.}{12}{29}{1999}
\referarticle{coestu00}{B. Coecke and I. Stubbe}{Int.~J.~Theor.~Phys.}{39}{601}{2000} 
\referarticle{eilML}{S. Eilenberg and S. Mac Lane}{Trans.~AMS}{58}{231}{1945}
\referarticle{EPR}{A. Einstein, B. Podolsky, and N. Rosen}{Phys.~Rev.}{47}{777}{1935}
\referarticle{faufro93}{Cl.-A. Faure and A. Fr\"olicher}{Geom.~Ded.}{47}{25}{1993}
\referarticle{faufro94}{Cl.-A. Faure and A. Fr\"olicher}{Geom.~Ded.}{53}{237}{1994}
\referarticle{faufro98}{Cl.-A. Faure and A.
Fr\"olicher}{Appl.~Cat.~Struc.}{6}{87}{1996}   
\referarticle{faumoo}{Cl.-A. Faure, D.J. Moore and C.
Piron}{Helv.~Phys.~Acta}{68}{150}{1995}
\referarticle{jaupir}{J.M. Jauch and C. Piron}{Helv.~Phys.~Acta}{42}{842}{1969}
\referbook{ML71}{S. Mac Lane}{Categories for the Working Mathematician}{}{Springer-Verlag, Berlin, 1971}
\referarticle{moo95}{D.J. Moore}{Helv.~Phys.~Acta}{68}{658}{1995}
\referarticle{moo97}{D.J. Moore}{Int.~J.~Theor.~Phys.}{36}{2707}{1997}
\referarticle{moo99}{D.J. Moore}{Stud.~Hist.~Phil.~Mod.~Phys.}{30}{61}{1999}
\referarticle{mul86}{C.J. Mulvey}{Rend.~Circ.~Math.~Palermo}{12}{99}{1986}
\referbook{pas00}{J. Paseka and J. Rosicky}{Quantales,}{in: Current Research in
Operational Quantum Logic, edited by B. Coecke, D.J. Moore and A. Wilce, pp.245--262}{Kluwer Academic
Publishers, Dordrecht, 2000}
\referarticle{pir64}{C. Piron}{Helv.~Phys.~Acta}{37}{439}{1964}
\referbook{pir76}{C. Piron}{Foundations of Quantum Physics}{}{W.A. Benjamin, Reading, 1976}
\referbook{pir98}{C. Piron}{M\'ecanique quantique. Bases et applications}{}{Presses
polytechniques et universitaires romandes, Lausanne, 1990 $\&$ 1998}
\referbook{ros90}{K.I. Rosenthal}{Quantales and their Applications,}{Pitman Research Notes
in Mathematics Series {\bf 234}}{Longman Scientific \& Technical, Essex, 1990}
\referbook{ros96}{K.I. Rosenthal}{The Theory of Quantaloids,}{Pitman Research Notes in
Mathematics Series {\bf 348}}{Longman Scientific \& Technical, Essex, 1996}
\referbook{stu}{I. Stubbe}{Notes du s\'eminaire itin\'erant des cat\'egories}{}{Universit\'e
de Picardie--Jules Verne, Amiens, 2000}
\referarticle{sou}{S. Sourbron}{Found.~Phys.~Lett.}{13}{357}{2000}
\end{itemize}  
\end{document}